\documentclass[
    ,final            % use final for the camera ready runs
  ]
  {aipproc}

\layoutstyle{6x9}

\begin{document}

\title{QCD corrections to neutralino annihilation}


\author{Heather E.\ Logan}{
  address={Ottawa-Carleton Institute for Physics, Carleton University, 
Ottawa K1S 5B6 Canada}
}

\begin{abstract}
We report on two recent calculations of QCD corrections to neutralino 
annihilation cross sections: (1) the next-to-leading order corrections 
to $\chi \chi \to gg$, and (2) the contribution to the cross section
for $\chi \chi \to q \bar q g$ arising from interference between the 
tree-level and loop-induced processes.  
\end{abstract}

\maketitle

%%%%%%%%%%%%%%%%%%%%%%%%%%%%%%%%%%%%%%%%%%%%
%% MAINMATTER
%%%%%%%%%%%%%%%%%%%%%%%%%%%%%%%%%%%%%%%%%%%%

\noindent
The presence of non-baryonic dark matter in the universe is compelling
evidence for physics beyond the Standard Model.  
Supersymmetry (SUSY) provides an especially attractive explanation with 
the lightest supersymmetric particle, usually a neutralino $\chi$,
as the dark matter candidate.
In this talk we review two recent calculations~\cite{dm1,dm2} of QCD 
corrections to neutralino annihilation via internal squark exchange.

The tree-level neutralino annihilation process $\chi\chi \to q \bar q$ 
via exchange of a squark of mass $M_{\widetilde q}$
can be reduced to an effective vertex described by a dimension-six operator
suppressed by $M_{\widetilde q}^2$,
\begin{equation}
  \mathcal{L}=(c/M_{\widetilde q}^2) \mathcal{O}_6~,
  \hspace{1cm} 
  \mathcal{O}_6=(\bar \chi \gamma_\mu \gamma_5 \chi )
   (\bar q \gamma^\mu \gamma_5 q).
\end{equation}
If the annihilation occurs at rest, the Majorana neutralino spinors reduce to 
a spin-singlet combination,
$\bar \chi \gamma_\mu \gamma_5 \chi \rightarrow 
K_\mu \bar \chi (\gamma_5 /\sqrt{2} m_\chi) \chi$, 
where $K$ is the center-of-mass
momentum of the two-neutralino system.
The operator ${\cal O}_6$ can then be written as the 
divergence of the axial-vector current,
\begin{equation}
 {\cal O}_6 \to 
 \left[ \overline\chi \ (i\gamma_5 / \sqrt{2}m_\chi)  \chi \right]  \,  
 \left[\partial_\mu (\overline q \gamma^\mu\gamma_5  q) \right]\, .
\end{equation}
In the massless quark limit, $m_q=0$, the axial vector current is conserved 
at tree level, $\partial_\mu (\overline q \gamma^\mu\gamma_5  q)=0$, and
all tree-level dimension-six amplitudes vanish in this limit.  
This is the well known partially-conserved axial current (PCAC) condition.

The quark mass suppression can be lifted in two ways:
\begin{enumerate}

\item By going beyond leading order in $\alpha_s$ to include the correction 
to the dimension-six operator that involves the anomalous triangle diagram.
This is $\chi \chi \to g g$.  This process first appears at order $\alpha_s^2$,
and thus suffers from a large QCD scale uncertainty.  We compute the 
next-to-leading order (NLO) QCD corrections to this process~\cite{dm1}.

\item By going to dimension-eight by including hard gluon radiation.  This 
is $\chi \chi \to q \bar q g$.  The tree-level diagram interferes with 
$\chi \chi \to g g^*$ with the off-shell gluon splitting into a $q \bar q$ 
pair, yielding a contribution to the cross section of order $\alpha_s^2$, the
same order as the leading $\chi\chi \to gg$ cross section.  We compute this
interference term~\cite{dm2}.
\end{enumerate}

%\section{$\mathbf{\chi \chi \to gg}$ at NLO}

\noindent
\underline{\bf $\mathbf{\chi \chi \to gg}$ at NLO:}
The leading-order (LO) amplitude for $\chi\chi \to gg$ in the massless quark 
limit is given by
\begin{equation}
    {\cal M}_{\rm loop} = \frac{\alpha_s}{2\sqrt{2} \pi}
    \sum_{q^{\prime}} \left[ \frac{|g_r|^2}{M_{\widetilde q_R^{\prime}}^2}
      + \frac{|g_{\ell}|^2}{M_{\widetilde q_L^{\prime}}^2} \right]
    i \epsilon^{\mu\nu\alpha\beta} \epsilon^{*a}_{\mu}(k_1) 
    \epsilon^{*a}_{\nu}(k_2) k_{1\alpha} k_{2\beta},
    \label{eq:Mloop}
\end{equation}
where the sum runs over the five light (massless) quark flavors $q^{\prime}$
and we neglect terms subleading in $m_{\chi}^2/M_{\widetilde q}^2$.
The $\chi q^{\prime} \widetilde q^{\prime}_{R,L}$ 
couplings $g_{r,\ell}$ for right and left quark helicities, respectively, are
\begin{equation}
  g_r = -\sqrt 2 N_{11} g^{\prime} Q, \hspace{0.5in} 
  g_{\ell} = - \sqrt 2 N_{11} g^{\prime} (T_3 - Q) + \sqrt 2 N_{12} g T_3.
  \label{eq:grgl}
\end{equation}
Here $T_3$ is the quark isospin, $Q$ is the quark electric charge, 
$g$ and $g^{\prime}$ are the weak couplings under $SU(2)_L$ and 
$U(1)_Y$, respectively, 
and $N_{11}$ and $N_{12}$ are the bino and wino components of the 
neutralino as defined in Ref.~\cite{DarkSUSY}. 
The matrix element in Eq.~(\ref{eq:Mloop}) results in an annihilation cross 
section for $\chi \chi \to gg$ of 
\begin{equation}
  v_{\rm rel} \sigma_{\rm dim6} = \frac{\alpha_s^2 m_{\chi}^2}{32 \pi^3} 
  \left\{ \sum_{q^{\prime}} 
  \left[ \frac{|g_r|^2}{M_{\widetilde q^{\prime}_R}^2}
    + \frac{|g_{\ell}|^2}{M_{\widetilde q^{\prime}_L}^2} \right] \right\}^2.
\label{eq:dim6xsec}
\end{equation}
For 6~GeV~$< m_{\chi} < 110$~GeV, this approximate formula is within 10\% 
of the result including the full quark mass dependence~\cite{dm1}.

In order to compute the NLO QCD corrections to
$\chi\chi \to gg$, we exploit the anomaly equation to relate this process
to pseudoscalar decay to gluons.
The conservation of the axial vector current is violated by the quark mass and
by the anomalous triangle diagram, in the form~\cite{ref:adlera}
\begin{equation}
  \partial_\mu (\overline q \gamma^\mu\gamma_5 q) = 
   2m_q \overline{q} i\gamma_5 q 
  + \frac{\alpha_s}{4\pi} G^{(a)}_{\mu\nu} \widetilde{G}^{(a)\mu\nu},
  \label{eq:anomaly}
\end{equation}
with 
$\frac{1}{2}\widetilde{G}_{\mu\nu} 
= \epsilon_{\mu\nu\alpha\beta} G^{\alpha\beta}$ 
denoting the dual color field strength tensor.  

The amplitude for $\chi \chi \to gg$ is obtained by setting $m_q = 0$ 
and neglecting the first term on the right hand side of 
Eq.~(\ref{eq:anomaly}).
The gluonic decay amplitude of a fundamental pseudoscalar, $A^0 \to gg$,
is also related to the anomaly equation.  This decay proceeds through
a quark loop; if the Yukawa coupling of this quark to $A^0$ is proportional
to the quark mass, the decay is dominated by the contribution of the 
heaviest quark in the loop.  In the heavy quark limit, $m_Q \gg m_A$,
we can neglect the divergence term on the left-hand side of 
Eq.~(\ref{eq:anomaly}).
Thus we see that two seemingly different processes, 
$A^0 \to gg$ through a heavy
quark loop and $\chi\chi \to gg$ through a light quark loop, are related
by the anomaly equation to the \emph{same} gluonic operator.
The Adler-Bardeen theorem \cite{ref:abj} guarantees that 
Eq.~(\ref{eq:anomaly}) is valid to all orders in $\alpha_s$.
We take advantage of this to obtain the NLO
QCD corrections to $\chi\chi \to gg$ from the known results for
$A^0 \to gg$ at NLO.

In the heavy top quark limit, for which our anomaly relation is
valid, the NLO QCD corrections to the $A^0 \to gg$ partial width 
are given by a
multiplicative factor times the LO decay rate~\cite{Spira:1995rr}.
Invoking the Adler-Bardeen theorem,
the NLO cross section for $\chi\chi \to gg$ in the zero-velocity limit
becomes~\cite{dm1}
\begin{equation}
  v_{\rm rel}  \sigma_{\rm NLO}(\chi \chi \to gg) = 
  v_{\rm rel} \sigma_{\rm dim6} 
  \left[1+\frac{\alpha_s}{\pi} \left( \frac{97}{4} - \frac{7}{6}N_f 
  + \frac{33 - 2 N_f}{6} \log \frac{\mu^2}{4 m_{\chi}^2}\right)
  \right],
\end{equation}
where $\mu$ is the renormalization scale.  
The strong coupling is evaluated based on $N_f$-flavor running 
at the scale $\mu$ where it appears both explicitly in the correction term and
within the leading-order cross section $v_{\rm rel} \sigma_{\rm dim6}$.
The integer $N_f$ counts 
the number of quark flavors in the gluon splitting, with $N_f = 5$ 
for $m_b \ll m_{\chi} \ll m_t$.  
Taking $\mu = 2 m_{\chi}$, $N_f = 5$, and $m_{\chi} = 100$~GeV, we obtain
$v_{\rm rel} \sigma_{\rm NLO}(\chi \chi \to gg) \simeq 
v_{\rm rel} \sigma_{\rm dim6} [1 + 0.62]$.
The NLO corrections reduce the residual renormalization scale 
dependence of the $\chi\chi \to gg$ annihilation cross section 
from $\pm 16\%$ to $\pm 9\%$ when we vary $\mu$ by a factor of two in 
either direction~\cite{dm1}.

%\section{$\mathbf{\chi \chi \to q \bar q g}$ interference term}
\vspace*{0.25cm}

\noindent
\underline{\bf $\mathbf{\chi \chi \to q \bar q g}$ interference term:}
To leading order in $m_{\chi}^2/M_{\widetilde q}^2$, it is
straightforward to extend the loop amplitude for $\chi\chi \to gg$ to include
one off-shell gluon splitting into a quark-antiquark pair:
\begin{equation}
    {\cal M}_{\rm split} = 
    - \frac{g_s}{\sqrt{2}} 
    \sum_{q^{\prime}} \left[ \frac{|g_r|^2}{M_{\widetilde q_R^{\prime}}^2}
      + \frac{|g_{\ell}|^2}{M_{\widetilde q_L^{\prime}}^2} \right]
    i \epsilon^{\mu\nu\alpha\beta} \epsilon^{*c}_{\mu} q_{3\alpha}
    (q_1 + q_2)_{\beta}
    \frac{\alpha_s \, \bar u(q_1) T^c \gamma_{\nu} v(q_2)}
         {2 \pi (q_1 + q_2)^2},
\label{eq:dim6}
\end{equation}
where $q_1$, $q_2$, and $q_3$ are the momenta of the final-state quark,
antiquark, and gluon, respectively, and the sum runs over the five 
light (massless) quarks in the loop.

The matrix element for the tree-level process $\chi \chi \to q \bar q g$ is
\begin{eqnarray}
  \mathcal{M}_{\rm tree} &=& \frac{g_s}{\sqrt{2}} 
  i \epsilon^{\mu\nu\alpha\beta} \epsilon^{*c}_{\mu} q_{3\alpha}
  (q_1 + q_2)_{\beta}
  \left\{ \frac{1}{2} \left[ \frac{|g_r|^2}{M_{\widetilde q_R}^4}
    + \frac{|g_{\ell}|^2}{M_{\widetilde q_L}^4} \right]
  \bar u(q_1) T^c \gamma_{\nu} v(q_2)
  \right. \nonumber \\ && \left.
  + \frac{1}{2} \left[ \frac{|g_r|^2}{M_{\widetilde q_R}^4}
    - \frac{|g_{\ell}|^2}{M_{\widetilde g_L}^4} \right]
  \bar u(q_1) T^c \gamma_{\nu} \gamma_5 v(q_2) \right\},
\end{eqnarray}
where we neglect terms subleading in $m_{\chi}^2/M_{\widetilde q}^2$.
Only the vectorlike part of $\mathcal{M}_{\rm tree}$ will interfere
with the dimension-six amplitude.  Summing over final-state polarizations
and colors and averaging over initial-state polarizations, we obtain
the interference term~\cite{dm2},
\begin{equation}
  \frac{1}{4} \sum_{\rm pols} 2 \, {\rm Re} \left[ \mathcal{M}_{\rm tree} 
  \mathcal{M}_{\rm split}^* \right]
  = - 4 \alpha_s^2 
  \sum_{q^{\prime}} \left[ \frac{|g_r|^2}{M_{\widetilde q^{\prime}_R}^2}
    + \frac{|g_{\ell}|^2}{M_{\widetilde q^{\prime}_L}^2} \right]
  \sum_q \left[ \frac{|g_r|^2}{M_{\widetilde q_R}^4}
    + \frac{|g_{\ell}|^2}{M_{\widetilde q_L}^4} \right]
  \left[ (q_1 \cdot q_3)^2 + (q_2 \cdot q_3)^2 \right],
\end{equation}
where we have summed over the five light internal quarks $q^{\prime}$ in 
$\mathcal{M}_{\rm split}$ of Eq.~(\ref{eq:dim6}) and over the five light 
external quarks $q$.
Integrating over the phase space, we find the contribution to the cross 
section from the interference term~\cite{dm2},
\begin{equation}
  v_{\rm rel} \sigma_{\rm int} = - \frac{\alpha_s^2 m_{\chi}^2}{32 \pi^3} 
  \frac{2 m_{\chi}^2}{3}
  \sum_{q^{\prime}} \left[ \frac{|g_r|^2}{M_{\widetilde q^{\prime}_R}^2}
    + \frac{|g_{\ell}|^2}{M_{\widetilde q^{\prime}_L}^2} \right]
  \sum_q \left[ \frac{|g_r|^2}{M_{\widetilde q_R}^4}
    + \frac{|g_{\ell}|^2}{M_{\widetilde q_L}^4} \right].
\label{eq:intresult}
\end{equation}
The form of this interference term is rather similar to that of
the dimension-six cross section given in Eq.~(\ref{eq:dim6xsec}).
In the special case of degenerate right- and left-handed squarks,
$M_{\widetilde q_R} = M_{\widetilde q_L} \equiv M_{\widetilde q}$,
we find that the interference term is related to the dimension-six cross
section by a multiplicative constant:
$v_{\rm rel} \sigma_{\rm int} = [ - 2 m_{\chi}^2 / 3 M_{\widetilde q}^2 ] 
v_{\rm rel} \sigma_{\rm dim6}$.
Combining this result with the NLO corrections to $\chi\chi \to gg$ with  
$N_f = 5$ light flavors and a renormalization scale $\mu = 2 m_{\chi}$, 
we obtain
\begin{equation}
  v_{\rm rel}\sigma = v_{\rm rel} \sigma_{\rm dim6}
  \left[ 1 + \frac{221}{12} \frac{\alpha_s^{(5)}(2 m_{\chi})}{\pi}
    - \frac{2}{3} \frac{m_{\chi}^2}{M_{\widetilde q}^2} \right].
\label{eq:combined}
\end{equation}
For $m_{\chi} \sim 100$ GeV the NLO correction is roughly 60\%; for, e.g., 
$M_{\widetilde q} \sim 2 m_{\chi}$, the interference term cancels off
about one quarter of the NLO correction.
We note that the expansion of the squark propagator in $\chi\chi \to gg$
yields correction terms starting at order $m_{\chi}^4/M_{\widetilde q}^4$
in the square brackets in Eq.~(\ref{eq:combined}); these are 
beyond our present interest.

%\section{Outlook}
\vspace*{0.25cm}

\noindent
\underline{\bf Outlook:}
In the region of SUSY parameter space where virtual squark exchange gives
a sizable contribution to neutralino annihilation, the neutralinos are 
``p-wave annihilators'' during freeze-out in the early universe; i.e., they
annihilate through the velocity-dependent $\chi\chi \to q \bar q$ process,
which is not suppressed by the quark mass.  In this situation,
the process $\chi\chi \to gg$ is subdominant and our QCD corrections
constitute only about $1-2\%$ of the dominant cross section.
They are thus of little importance for computing the relic neutralino 
abundance.  However, for neutralino annihilation at the present time 
the relative neutralino
velocity is much lower, leading to a much smaller tree-level $\chi\chi\to
q \bar q$ cross section.  In this situation, the $\chi\chi \to gg$ cross
section can be as large or larger than $\chi\chi \to q \bar q$, so that
the QCD corrections computed here could have a significant impact on the 
computation of gamma ray and neutrino fluxes from neutralino annihilation 
in the galatic halo and inside the sun, respectively.
In~\cite{dm2} we provided the spin-averaged squares of
matrix elements for the dimension-six, dimension-eight, and interference
term contributions to $\chi\chi \to q \bar q g$ in terms of the final-state
particle momenta to facilitate their implementation into
Monte-Carlo generators for the computation of indirect dark matter 
detection rates.

%%%%%%%%%%%%%%%%%%%%%%%%%%%%%%%%%%%%%%%%%%%%%%%%
%% BACKMATTER
%%%%%%%%%%%%%%%%%%%%%%%%%%%%%%%%%%%%%%%%%%%%%%%%

\vspace*{0.25cm}

%\begin{theacknowledgments}
I thank Vernon Barger, Wai-Yee Keung, Gabe Shaughnessy, and 
Adam Tregre for an enjoyable and fruitful collaboration that led to these
results.  
%I also thank the organizers of SUSY'06, especially Jonathan Feng,
%for putting together such a wonderfully stimulating conference.
This work was supported by the Natural Sciences and Engineering Research
Council of Canada.
%\end{theacknowledgments}



\bibliographystyle{aipproc}   % if natbib is available

%\bibliography{sample}

%%%%%%%%%%%%%%%%%%%%%%%%%%%%%%%%%%%%%%%%%%%
%% The following lines show an example how to produce a bibliography
%% without the help of the BibTeX program. This could be used instead
%% of the above.
%%%%%%%%%%%%%%%%%%%%%%%%%%%%%%%%%%%%%%%%%%%

\end{document}